\documentclass[twocolumn,superscriptaddress,nofootinbib,prl]{revtex4-1}
\usepackage{epsfig}
\usepackage{amssymb}
\usepackage{amsmath}
\usepackage{amsbsy}
\usepackage{color}
\usepackage{ulem}
\usepackage{epstopdf}
\begin{document}
\title{Topological free volume and quasi-glassy dynamics in melt of ring polymers}
\date{}
\author{Takahiro Sakaue}
\email{sakaue@phys.aoyama.ac.jp}
\affiliation{Department of Physics and Mathematics, Aoyama Gakuin University, 5-10-1 Fuchinobe, Chuo-ku, Sagamihara, Kanagawa 252-5258, Japan}
\affiliation{PRESTO, Japan Science and Technology Agency (JST), 4-1-8 Honcho Kawaguchi, Saitama 332-0012, Japan}

\begin{abstract}
Motivated by recent observations that non-concatenated ring polymers in their dense solution exhibit a glass-like dynamics, we propose a free volume description of the motion of such rings based on the notion of topological volume.
We first construct a phenomenological free energy which enables one to qnaitify the degree of topological crowding measured by the coordination number. Then we pinpoint a key role of the cooperative dynamics of neighboring rings, which is responsible for an anomalous dependence of the global structural relaxation (diffusion) time on ring length. Predictions on molecular weight dependence of both static (ring size, coordination number) and dynamic (relaxation time, diffusion coefficient) quantities are in very good agreement with reported numerical simulations.
Throughout the discussion, the entanglement length $N_e$ is assumed to be a unique characteristic length for the topological constraint, hence, all the physical quantities are universally described in terms of the rescaled chain length $N/N_e$. Finally, we discuss how the dense solution of rings is analogous yet different from ordinary glassy systems.
\end{abstract}


\maketitle

\section{Introduction}
Despite its ubiquity in biology and potential applicability in material science, the behavior of ring polymers remains largely mysterious in many respects, understanding of which lags far behind the linear polymer counterpart~\cite{McLeish2005,McLeish2008}. The main source of difficulty arises from the inevitable constraint that a topological state on inter-ring concatenation and intra-ring knotting  has to be rigorously conserved at any later stage unless the bond breakage occurs.
Recent experiments and simulations have raised several puzzles in dynamics of dense solution of non-concatenated rings~\cite{Richter_2014,Michieletto_2016,Michieletto_2017}, hinting some analogy to the glass transition~\cite{Berthier_2011,Liu_2010}. The conceived state, dubbed as topological glass, however seems to be very different from ordinary glass in that the large scale dynamical anomaly entails essentially no motional restriction at the scale of constituents (monomers).
Here, we propose a viewpoint that qualifies rings as ultra-soft particles, whose effective volume arises from the topological constraint (TC). In dilute solution, two rings in unlinked topology feel an entropic repulsion upon close approach, since the unlinking (non-concatenation) TC reduces the number of available conformations (Fig.~\ref{Fig1}(a)). In dense solution of unlinked rings, we are naturally led to think of the free volume of the effective volume of topological origin; as we shall see this {\it topological free volume} is intimately connected to the coordination number of rings. Bringing these concepts together, we set up a stage for statistical mechanical analysis, which yields various predictions in remarkable agreement with reported observations. 
It unveils a rational scenario through which the large scale cooperative dynamics of rings emerges; the physical picture indeed bears some similarity with ordinary glassy dynamics, but with a cardinal difference originating from the ultra soft nature of rings.

\section{Topological constraint in linear vs. ring polymers}
To make the nature of TC in ring polymer system clear, let us recall the phenomenon of entanglement in linear polymer melt, which is also a consequence of TC. 
Consider a melt of flexible polymers, each of which consists of $N$ monomers of size $a$. Because of the screening of excluded-volume interactions, the size of individual linear polymers is $R \sim a N^{1/2}$, hence the number of overlapping polymers with any reference polymer defined as
\begin{eqnarray}
X \sim R^3/(N a^3) 
\label{X_define}
\end{eqnarray}
 is evaluated as $X \sim  N^{1/2}$~\cite{Rubinstein_Colby}.
Consequence is that the entanglement is inevitable in dense solution of long polymers. The modern development of rheology of entangled polymers is based on the geometrical picture, where a polymer is confined in a virtual tube made up of the surrounding polymers. The reptation model describes the motion of polymer in the tube, where the chain ends play a pivotal role, emphasizing the transient (though long-lived) nature of the TC~\cite{de_Gennes_scaling, Rubinstein_Colby}.

\begin{figure*}[t]
\begin{center}
\includegraphics[width=0.95\textwidth]{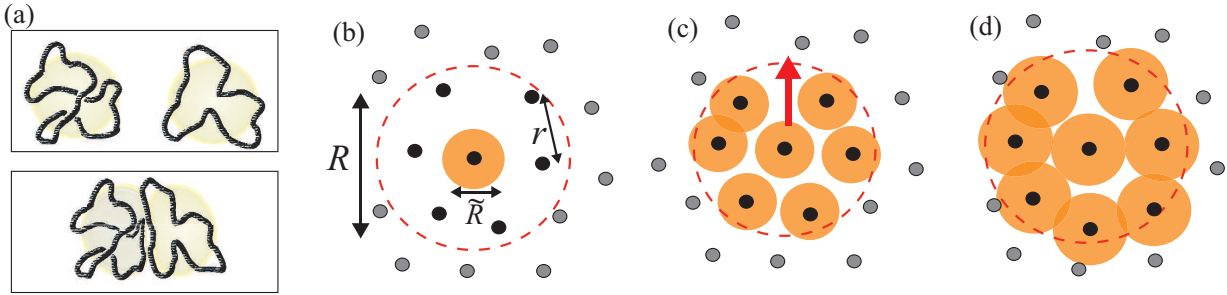}
\caption{Schematics of the system of non-concatenated rings. (a) For two rings close to each other (bottom), the number of their possible configurations is reduced from the free state (top) due to the non-concatenation TC. This produces an entropic repulsion between them~\cite{Frank-Kamenetskii1975} .  (b) Length scales in the system. A central dot is the center of mass (COM) of a reference ring, whose spatial size $R$ is indicated by a dotted circle. Other black dots inside the dotted circle denote the COMs of rings overlapping with the reference ring, which defines the number of overlapping rings $X$, i.e.,{\it coordination number}. A shaded circle represents the effective size ${\tilde R}$ of a ring due to the TC. In a particle picture, the average distance $r$ between COMs of neighboring rings is a key quantity to control the global structural relaxation. In contrast to (c) short ring case $N < N_e$, (d) TC for longer ring $N > N_e$ imposes a severe crowding (larger $X$, thus, higher topological volume fraction $\Psi$) in such a way that any ring is caged by its surroundings. As in ordinary crowded systems, the global structural relaxation requires the cage breaking realized by the passage of a ring through its ''gate", but the solitary motion of any single ring is ineffective.However, the gate opening could be achieved by a cooperative motion of $M(N)$ neighboring rings.  
 }
\label{Fig1}
\end{center}
\end{figure*}

\begin{figure}[h]
\begin{center}
\includegraphics[width=0.39\textwidth]{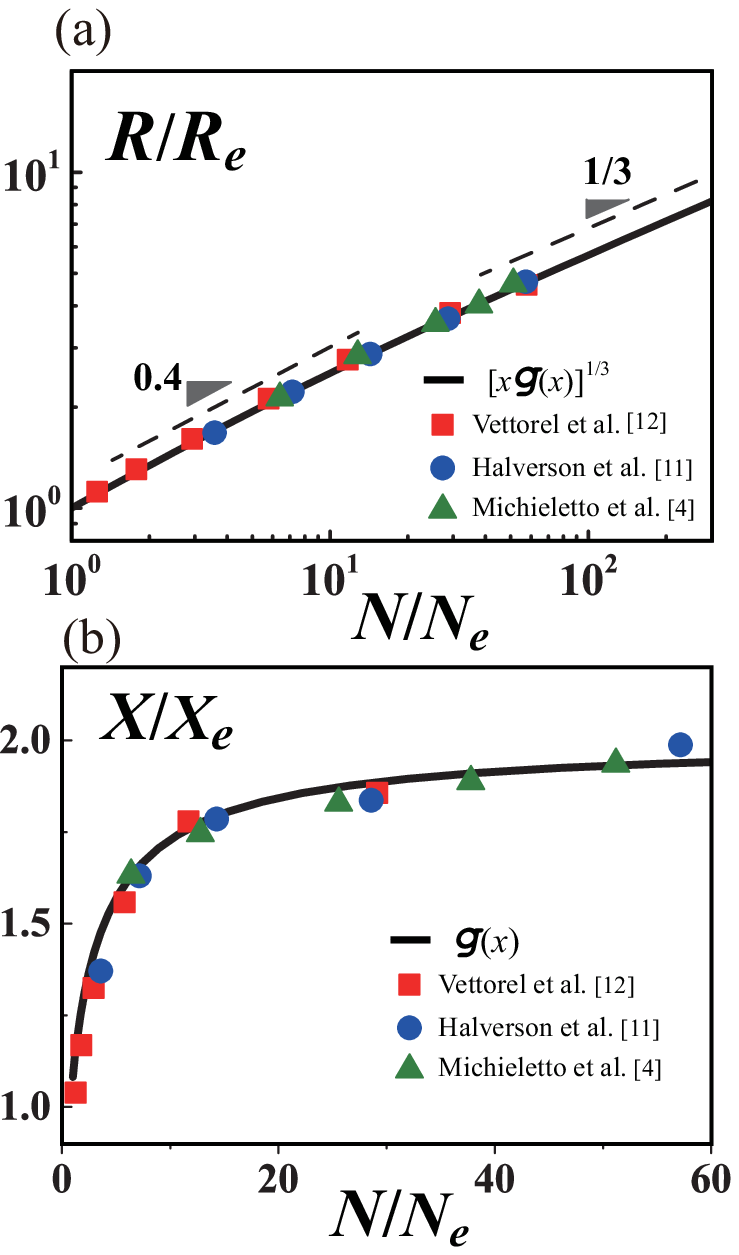}
\caption{Comparison of free energy prediction (solid line) with numerical simulation data for spatial size $R$ and coordination number $X$ of rings in their non-concatenated melt of dense solution. As a measure of ring size, we adopt here the spanning distance $R \equiv R^{(s)}$. Numerical data are obtained from literature  (squares~\cite{Vettorel_2009}, circles~\cite{Halverson_2011}, triangles~\cite{Michieletto_2016})  (a) Plot of $R/R_e$ as a function of $N/N_e$. Before reaching the asymptotic compact scaling with $\nu=1/3$, there exits a broad crossover up to $N/N_e \rightarrow N_c/N_e \sim 30$ (see Appendix for the estimation of $N_c$). (b) Plot of $X/X_e$ as a function of $N/N_e$. See Table 1 for details of simulation models and the estimated values of $R_e$ and $X_e$.
}
\label{Fig2}
\end{center}
\end{figure}
In the ring polymer counterpart, there are now a large body of indications that the size is much more compact characterized by the smaller exponent; $R \sim N^{\nu}$ with $\nu =1/3$ for large $N$, but with a prefactor substantially larger than $a$, i.e., $R/(a N^{1/3}) \gg 1$~\cite{Halverson_2011,Vettorel_2009,Suzuki_2009,Sakaue_2011,Sakaue_2012,Rosa_2014,Obukhov_2014}. This results in $X \sim const. (\gg 1)$ indicative of a qualitatively different TC from classical entanglement picture of heavily overlapping threads.
Rather it implies a soft particle-like behavior, which allows some degree of overlapping. 
Indeed, as might be intuitively clear, the TC endows any reference ring with entropic repulsion against surrounding rings which are unlinked to it~\cite{Frank-Kamenetskii1975} (Fig.~\ref{Fig1} (a)).
One can then define the length 
\begin{eqnarray}
r \propto R/X^{1/3}, 
\label{r_ring}
\end{eqnarray}
which measures the average distance between neighboring rings (Fig.~\ref{Fig1} (b)). This is analogous to the inter-particle distance $r \propto b/\psi^{1/3}$ in the system of rigid particles (with size $b$ and the particle volume fraction $\psi$). Such a consideration suggests an interesting possibility to investigate the problem of dense ring polymers under TC with the concepts developed in the liquid and glass transition physics~\cite{Berthier_2011,Liu_2010}.
Unlike the rigid particle case $r > b$, however, one admits here a strong overlapping $r < R$ reflecting ultra-soft nature of the ring (Fig.~\ref{Fig1}). 
Note that numerical and experimental observations suggest that the exponent $\nu=1/3$ is realized only for very long rings, and many of the practical cases may fall into a broad crossover characterized by some effective exponent bounded as $ 1/3 < \nu_{eff} < 1/2$.

\section{Free energy in terms of the topological volume fraction}
\label{TC_free_energy}
The primal control parameter in the problem is the ring length $N$, which plays the similar role as the temperature $T$ in the ordinary glass transition phenomenology. 
For the systematic study, one therefore needs a framework, which describes how $R$ and $X$ depends on $N$. 
Here, we utilize the phenomenological free energy, which is constructed based on the following two observations; (i) as rings behave like soft particles due to TC, there should be associated {\it topological volume}, and (ii) the TC in rings becomes relevant above some characteristic length scale $N_e$~\cite{Halverson_2014, Ge_2016}, which implies the ideal chain statistics in smaller scale, thus 
\begin{eqnarray}
X_e \equiv X(N=N_e)  \sim R_e^3/(N_e a^3) \sim N_e^{1/2}
\label{Xe_define}
\end{eqnarray}
, where $R_e \sim a N_e^{1/2}$ is the spatial extent of the $N_e$-strand.
Note that the second observation is akin to an empirical fact admitted in the entanglement effect in linear polymers~\cite{de_Gennes_scaling, Rubinstein_Colby}, which is the reason we employed the symbol $N_e$ for the characteristic chain length.
The simplest free energy in accordance with the above observations is~\cite{Sakaue_2011,Sakaue_2012,Sakaue_2016}
\begin{eqnarray}
\frac{F}{k_BT} =  - \ln{\left(1-\frac{X}{X_c}\right)} + \frac{N}{N_e} \left( \frac{X_e}{X}\right)^{1/(3\nu_{0}-1)}
\label{F_total}
\end{eqnarray}
where $X_c$ represents the maximum number of overlapping rings achieved in large $N$ limit, and $\nu_{0} \simeq 0.588 $ is a critical exponent describing the size of self-avoiding chain (see Appendix for more discussion on the proposed free energy).
The quantity $\Psi \equiv X/X_c$ may be regarded as a {\it topological volume fraction}. To motivate such a usage, let us rewrite Eq.~(\ref{r_ring}) for the inter-ring distance as
\begin{eqnarray}
r = \left( \frac{R}{X_c^{1/3}}\right)\left( \frac{X}{X_c}\right)^{-1/3} = {\tilde R} \Psi^{-1/3}
\label{R_eff}
\end{eqnarray}
Now $\Psi$ is bounded as $\Psi \le 1$ by definition, and the comparison with rigid particle case qualifies ${\tilde R} = RX_c^{-1/3}$ and $\Psi$ as the effective size and corresponding volume fraction associated with TC.
The first term in Eq.~(\ref{F_total}) takes a form from celebrated van der Waals theory, which represents the reduction in the free volume, in the present context, reflecting the entropic repulsion due to non-concatenation TC.
This term favors the smaller $\Psi$, thus leads to the ring shrinkage. Counteract to this is the second term, which represent an entropic penalty associated with squeezing a ring while keeping the unknotting constraint, and can be derived from standard scaling argument (see Appendix).

The free energy ~(\ref{F_total}) has a desired property $F \sim k_BT$ at $N = N_e$ with a supplement $\Psi_e \equiv X_e/X_c \simeq 0.5$, ensuring the onset of TC at this length scale in accordance with the above observation (ii).
For longer rings, the optimum $X$ as a function of $N$ can be derived by minimizing Eq.~(\ref{F_total}). We are thus led to a generic form $X(N) = X_e \  {\mathcal G}(N/N_e)$ with a function ${\mathcal G}(x) \rightarrow 1$ at $x=1$ and ${\mathcal G}(x) \rightarrow X_c/X_e$ at $x \gg 1$. From this, the ring size follows as $R(N) \simeq R_e (N/N_e)^{1/3}[{\mathcal G}(N/N_e)]^{1/3}$. 
In Fig.~\ref{Fig2}, we compare predictions from our free energy (with $\Psi_e=0.5$) with three set of numerical simulation data. Here we adopt the spanning distance $R^{(s)}$ (root mean-square distance between monomers $N/2$ apart) as a measure of the spatial size of rings, so that $X$ is the average number of surrounding rings, whose center of mass (COM) are located within the distance $R^{(s)}$ from the COM of reference ring. 
Note that the values $R_e$ and $X_e$ are evaluated from the numerical data in references~\cite{Vettorel_2009, Halverson_2011, Michieletto_2016} (see Table 1), which play a role of rescaling factors in the master plot.
While the resultant $R_e$ values reasonably accords with theoretical expectation (Appendix), it is striking that all the data set point almost the same value for $X_e \simeq 8$ despite different simulation models and conditions employed. The same conclusion is reached if we measure the ring size with radius of gyration $R^{(g)}$, in which case $X_e \simeq 1.4$ is found (Appendix, Fig.~\ref{SFig1}).
Hence, besides the quantitative predictability of the proposed free energy, we find that the coordination number serves as a robust measure for TC. In particular, the onset of TC can be identified by $X_e$, whose precise value is independent of the system details. The crossover towards the compact statistic regime for longer rings are quantified by the function ${\mathcal G}(x)$, which describes a gradual increase of the coordination number from $X_e$ towards $X_c$. 

Here it is instructive to draw a connection of the particle-like behavior of non-concatenated rings in their dense solution with the argument based on the random packing.
For the assembly of particles to be mechanically stable, Maxwell's criterion claims that the minimum coordination number $X_{iso}=2 d_f$ is required, where $d_f$ is number of degrees of freedom determined by the symmetry of the constituent particles~\cite{Liu_2010}. With $d_f = 6$ for a generally shaped particle in $3$ dimension, we see an impressive consonance of $X_{iso}$ with $X_e$ and $X_c =X_e/\Psi_e$ when we adopt the spanning distance $R^{(s)}$ as a measure of ring size.
Note also that the range of the value $X \in (X_e, X_c)$ in this case corresponds to the so-called Kavassalis-Noolandi number to mark the onset of entanglement effect in dense linear polymer solutions~\cite{Kavassalis_1987, Rubinstein_Colby}. This suggests a fundamental role of $X$ to describe the TC in polymer systems. We also emphasize that the function ${\mathcal G}(x)$ is universal in the sense that it does not depend on the system parameters (Appendix).


\begingroup
\squeezetable
\begin{table}[h]
 \caption{A summary of methods and numerical values employed in or extracted from simulations ($l_p$: persistence length). The values $R_e^2$ and $X_e$ are obtained by inter-(or extra-)polating the data in references. All lengths are measured in unit of lattice size (MC) or bead diameter (MD).}
 \small
  \begin{tabular}{l||c|r|r|} \hline
      Reference & \cite{Vettorel_2009} & \cite{Halverson_2011} & \cite{Michieletto_2016} \\ \hline \hline
    Method & MC (cubic lattice) & MD & MD \\
    Stiffness & no angle potential & $l_p \simeq 1.5 \sim 2$ & $l_p \simeq 5 $ \\
    Volume fraction & 0.5 & $0.85 $ & $0.1 $ \\
    $N_e$ & 175 & 28 & 40  \\ 
    $(R_e^{(s)})^2$ & 75 & $18 $ & $88 $\\
    $X_e|_{R= R^{(s)}}$ & 7.7 & 8.4 & 7.8 \\ 
    $(R_e^{(g)})^2$ & 26 & $6.1$ & $30 $ \\
    $X_e|_{R= R^{(g)}}$ &n/a  & 1.4 & 1.3 \\ \hline
  \end{tabular}
 
\end{table}
\endgroup

\section{Cooperative mechanism for global structural relaxation}
To discuss the consequences of topological volume on ring dynamics in dense solution, let us recall a topological length $N=N_e$. It indicates Rouse dynamics for shorter scale,  thus introduce a time scale $\tau_e \sim \tau_0 N_e^2$ for the relaxation of $N_e$-strand (with $\tau_0$ being a monomeric time scale), which plays a role of an elementary time scale for larger scale dynamics, where TC matters.
Here, while recent rheological measurements have evidenced a self-similar stress relaxation process without sign of rubber plateau~\cite{Kapnistos_2008,Doi_2015}, neutron spin-echo experiment detected glassy dynamics of rings' COM~\cite{Richter_2014}. A connection between these two observations remain unsolved. A hint comes from a detailed analysis of numerical simulation data, which indicates that while the stress relaxation is correlated to the internal conformational reorganization, whose characteristic time scales as $\tau^{(1)} \simeq \tau_e (N/N_e)^{z_1}$ with $z_1 \simeq 2.2$, the diffusion would be much slower, that is to say, $\tau \gg \tau^{(1)}$ for large $N$, where $\tau$ is the time scale for global structural relaxation during which the ring travels its own size~\cite{Halverson_2011_2}. 
It is likely that a reason for such a self-diffusion slower than expected lies in a cooperative dynamics, which is necessary for the global structural relaxation. 

\begin{figure}[ht]
\begin{center}
\includegraphics[width=0.46\textwidth]{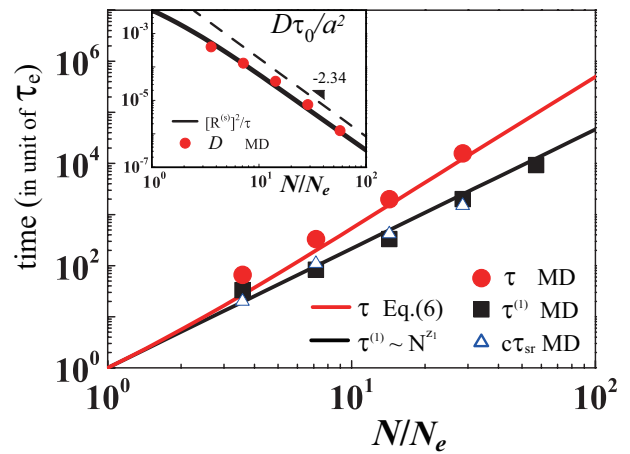}
\caption{Internal ($\tau^{(1)}$)  and global ($\tau$) structural relaxation times as a function of $N/N_e$ compared with MD simulation data~\cite{Halverson_2011_2} with $\tau_e = 120 \ \tau_0$. Also plotted is stress relaxation time $\tau_{sr}$ (multiplied by a factor $c=0.4$) evaluated in Ref.~\cite{Ge_2016} using the data in Ref.~\cite{Halverson_2011_2}. Here, we adopted $z_1 = 7/3$ suggested in ref.~\cite{Ge_2016}. (Inset) Diffusion coefficient. Here the line with $[R^{(s)}]^2/\tau$ is compared with the diffusion coefficient measured in MD simulation~\cite{Halverson_2011_2}. 
See also ref.~\cite{Jung_2015} for a very similar numerical data on the separation between $\tau^{(1)}$ and $\tau$. }
\label{Fig3}
\end{center}
\end{figure}

As a prototype, we begin with reviewing a minimal description of cooperative dynamics in glass transition~\cite{Salez_2015}.
Consider an assembly of particles of radius $b$ with volume fraction $\psi$ such that the average interparticle distance $r \propto b \psi^{-1/3}$. The mechanism for structural relaxation would depend on the average size $\Delta \sim r-2b$ of gate, through which each particle escapes from its cage. 
Two characteristic interparticle distances $r_c$ and $r_v$ can be conceived; while the latter corresponds to the kinetic arrest point $r_v \simeq 2b$, the former $r_c (> r_v)$ signals the onset of crowding effect so that, at higher concentration ($r < r_c$), the cage breaking requires the cooperative motion of adjacent particles.
Since the missing space for the gate opening is $r_c - r$, and the space created by cooperative motion of $M-1$ neighboring particles is $(M-1) \Delta$, it follows
\begin{eqnarray}
M(\psi) \sim \frac{r_c -r_v}{r-r_v} \sim \frac{(\psi_v/\psi_c)^{1/3}-1}{(\psi_v/\psi)^{1/3}-1}
\label{M_colloid}
\end{eqnarray}
where $\psi_c \propto (b/r_c)^3$, $\psi_v \propto (b/r_v)^3$ are volume fraction corresponding, respectively, to $r_c$, $r_v$.
From this, the structural relaxation time $\tau$ (in unit of molecular time scale $\tau_m$) is
\begin{eqnarray}
\frac{\tau}{\tau_m} \sim \left( \frac{\tau_{liq}}{\tau_m}\right)^{M(\psi)}
\label{tau_alpha}
\end{eqnarray}
where $\tau_{liq}$ is a typical liquid-like relaxation time at $\psi=\psi_c$, indicating a drastic slowing down of dynamics due to the crowding at $\psi>\psi_c$ towards sharp divergence of $\tau$ at $\psi=\psi_v$. To recast Eq.~(\ref{tau_alpha}) in the form of temperature dependence, one usually invokes the thermal expansion of materials, which is assumed to be described by $\psi(T) \simeq \psi_v[1 + \alpha(T_v - T)]$ in the range $(\psi_c, T_c)$ to $(\psi_v, T_v)$. Combining this with Eqs.~(\ref{M_colloid}) and~(\ref{tau_alpha}), one obtains the Vogel-Fulcher-Tammann (VFT) relation
\begin{eqnarray}
\tau \sim \tau_m  \exp{\left( \frac{A}{T-T_v}\right)}
\end{eqnarray}
with $A  = (T_c-T_v) \ln{(\tau_{liq}/\tau_m)}$.

Applying the above picture of the cooperative dynamics to the slow diffusion of rings under TC, the notion of topological volume invites a correspondence $\psi \rightarrow \Psi$. The cooperative onset $\psi_c$ then naturally corresponds to the TC onset $\Psi_e$ at $N_e$ with the replacement of characteristic time scales $\tau_m \rightarrow \tau_e$ and $\tau_{liq} \rightarrow \tau^{(1)}$. We then set $r_v \rightarrow 0$, which amouts to saying that there is no spontaneous kinetic arrect due to the ultra-softness of rings, i.e., their COMs can indeed overlap in space.
Such a translation leads to the growing cooperativity $M(N) \sim (\Psi(N)/\Psi_e)^{1/3} 
=[{\mathcal G}(N/N_e)]^{1/3}$ with $N$,
hence, the global structural relaxation time
\begin{eqnarray}
\tau \sim  \tau_e \left( \frac{\tau^{(1)}}{\tau_e}\right)^{M(N)} \sim   N^{z_{eff}}
\label{tau_N}
\end{eqnarray}
where the exponent $z_{eff} = z_1 \times  [{\mathcal G}(N/N_e)]^{1/3}$ increases from $z_1$ at $N_e$ towards $z_{\infty} = z_1 [X_c/X_e]^{1/3} \sim 3$
 in the long chain asymptote (Appendix Fig.~\ref{SFig2}).
Notice that the $\Psi - N$ relation (or equivalently $X - N$ relation shown in Fig.~\ref{Fig2}(b)) determined by our free energy plays an analogous role as the $\psi - T$ relation (thermal expansibility) in the context of the derivation of VFT relation in particle systems. 


As demonstrated in Fig.~\ref{Fig3}, Eq.~(\ref{tau_N}) captures all the essential features observed in numerical simulations.
In addition to semi-quantitative prediction for $\tau$ and its departure from $\tau^{(1)}$, it provides the diffusion coefficient $D(N) \sim R(N)^2/\tau(N)$, whose scaling with $N$ remarkably agrees with numerical observations~\cite{Michieletto_2016,Halverson_2011_2} (Fig.~\ref{Fig3}(inset); see also Appendix Fig.~\ref{SFig2}).
Compared to previous predictions based on the conventional estimation $D^{(1)} \sim R(N)^2/\tau^{(1)}(N)$~\cite{Ge_2016,Smrek_2015}, this improvement on $D$ once again set forth the notion of slow diffusion due to the cooperative structural relaxation.
Yet noticeable is the absence of VFT-like divergence in $\tau$ despite invoking mandatory cooperative motion for the cage breaking. In our description, this traces back to the ultra-soft nature of rings $r_v \rightarrow 0$, thus no divergence in the cooperativity $M$. Still, there is a small build up in $M$ at $N > N_e$, and combined with the slowing down in single ring dynamics $\tau^{(1)}$, it acts as a multiplicative factor in exponent, leading to unconventional, i.e., faster than power-law, increase in $\tau$ with $N$.

\section{Conclusions}
To summarize, we have proposed a cooperativity scenario for slow dynamics in dense solution of non-concatenated ring polymers motivated by some analogy between soft colloids and rings. 
While we have shown that many of puzzling observations could be resolved by introducing the notion of topological volume, it should be recognized as a minimal description in the sense that we only look at COM degrees of freedom of rings to evaluate the cooperativity $M$, which is mirrored in a small $M_c = (X_c/X_e)^{1/3} \sim 1.3$ in the large $N$ limit.
More realistic picture would be that a larger number of neighboring rings behave collectively not only by the COM translation but also by internal deformation modes to create a gate for the structural relaxation. 
Our discussion yields several challenges and future directions.
(i) Quantifying the degree and the nature of the cooperativity in ring solutions -- it should help a possible refinement of our current description.
(ii) Clarifying the relation with threading picture~\cite{Michieletto_2016, Tsalikis_2016} --  there seems to be some correspondence between our structural relaxation time $\tau$ and the ring dethreading time possibly responsible for the slow mode in stress relaxation~\cite{Tsalikis_2016}.
(iii) Elucidating the analogy with soft colloids -- they are known to be strong glass formers~\cite{Mattsson_2009}, and rings could be viewed as the softest ''particle" ever known. Compared, for instance, to star polymers~\cite{Likos_1998}, the energy scale of the effective repulsive pair potential is much lower (on the order of $k_BT$), and COMs of different rings can spatially overlap rather easily. A quantitative comparison with the dense solution of soft core (such as the Gaussian core) model may yield useful insights~\cite{Ikeda_2011}.
(iv)Apparently similar soft colloidal picture (with comparable energy scale) is known to describe the linear polymer systems as well~\cite{Louis_2000}. Unlike the ring polymer systems, however, it is nothing to do with the conformation of individual chains, and the correlation hole (or effective repulsion) becomes weaker with the chain length. Nontheless, some anomalous feature in the diffusive dynamics of COM~\cite{Richter_2014, Michieletto_2016, Halverson_2011_2}, which is not captured by the reptation theory, may be described by the cooperative mechanism scenario proposed here.  
(v) Building a microscopic foundation for the proposed description -- should be a fundamental problem. Some attempts for coarse-graining of ring polymer systems indicate the crucial effect of TC~\cite{Likos_2010, Likos_2014}.
(vi) Exploring a novel depletion effect~\cite{Asakura_Oosawa_1954} -- the notion of topological volume indicates an associated depletion effect. 
This would be related to the recent prediction on the promotion of phase separation in the blend of ring polymers with dissimilar molecular weights~\cite{Sakaue_2016,Doi_2018}, and may have non-trivial effects in confined spaces, e.g. in cell nucleus.

\section*{Acknowledgements}
This work was supported by KAKENHI (No. JP16H00804, ``Fluctuation and Structure") from MEXT, Japan, and JST, PRESTO (JPMJPR16N5).T.S. thanks D. Michieletto for enlightening discussion and providing numerical data in ref.~\cite{Michieletto_2016}.

\section*{Appendix}
\subsection*{Some discussion on free energy associated with TC}
\paragraph*{Entropic penalty due to non-concatenation constraint}
A unique feature in the system of unlinked rings compared to the standard particle system is that constituents have a freedom to adjust their spatial size to reduce the repulsive interaction (here originated from the non-concatenation TC among neighboring rings). In the free energy~(\ref{F_total}), it is accounted for through the van-der-Waals term $F_{unlink}/k_BT =- \ln{(1-\Psi)}$  based on the free volume concept.

Let $c$ denote the number concentration of rings in the melt. From free energy $F_{unlink}$ per ring, we obtain the free energy density $f_{unlink} =  c F_{unlink}$ as
\begin{eqnarray}
\frac{f_{unlink}}{k_BT} = - \frac{X}{R^3}\ln{\left( 1-\frac{X}{X_c}\right)} = B_2 c^2 + B_3 c^3 + \cdots 
\label{f_unlink}
\end{eqnarray}
where we have used the relation $c \equiv N_{ring}/\Omega = \phi/(Na^3)$ ($N_{ring}$ is the total number of rings in the system volume $\Omega$, and the monomer volume fraction $\phi \sim 1$ in the melt state), and the definition of the coordination number $X$ given as Eq.~(\ref{X_define}). The final expression is obtained by the expansion into virial series, which identifies the virial coefficients $B_2 \simeq R^3/X_c$, $B_3 \simeq R^6/2X_c^2$, etc. Note that the second virial coefficient is in line with the effective ring size ${\tilde R}$ invoked in Eq.~(\ref{R_eff}), i.e., $B_2 \simeq {\tilde R}^3$, that is, the soft nature of the ring is linked with $X_c = X_e/\Psi_e$, hence with $N_e$.
It is instructive to compare our free energy $F_{unlink}$ with that conjectured by Cates and Deutsch long ago. They suggested the form $F_{unlink}/k_BT \sim R^3 /(Na^3) \sim X$~\cite{Cates_Deutush_1986}. In light of the present discussion, their free energy corresponds to the second virial approximation (which amounts to setting $B_n =0$ for integer $n>2$) with neglecting the softness factor $X_c$ ($B_2 \sim R^3$). Therefore, it fails to capture the many-body effect responsible for the compact statistics ($\nu = 1/3$) in the long $N$ limit, and lacks $N_e$ as a unique characteristic length in the problem.

\paragraph*{Entropic penalty due to unknotting constraint}
The free energy $F_{unlink}(\Psi)$ due to the non-concatenation constraint favors the smaller $\Psi$, hence the ring shrinkage. 
More shrunk, however, another TC associated with unknotting within individual rings becomes more relevant. Here we briefly outline the derivation of the entropic penalty due to the unknotting TC (second term in Eq.~(\ref{F_total})) following the discussion in ref.~\cite{Sakaue_2016}~\footnote{In refs.~\cite{Sakaue_2011,Sakaue_2012}, this term has the form $(N/N_e)(X_e/X)^2$, which amounts to set $\nu_{0}=1/2$, but it was corrected later to account for the swelling of unknotted ring due to TC~\cite{Sakaue_2016}; see the discussion below.}.

According to the standard scaling argument, the entropic cost of squeezing a polymer into the size $R$ is evaluated as
\begin{eqnarray}
\frac{F_{unknot}}{k_BT} \sim \left( \frac{R_0}{R}\right)^{\beta}
\label{F_unknot_scaling}
\end{eqnarray}
where $R_0$ is the size of a ring in reference state~\cite{de_Gennes_scaling}. In the present problem, a ring in reference state can be prepared by hypothetically switching off the non-concatenation TC among different rings, i.e., it is free from ordinary excluded-volume effect (i.e., screened in melt), and not perturbed by the non-concatenation TC, but constrained to keep the unknot topology. Such a ring is essentially identical to an ideal, but non-phantom unknot ring in dilute solution, whose size is given by
\begin{eqnarray}
R_0 \sim \left\{
\begin{array}{ll}
a N^{1/2} & \quad (N < N_0) \\
a N_0^{1/2} (N/N_0)^{\nu_0}  & \quad (N \gg N_0)
\end{array}
\right.
\end{eqnarray}
where $N_0$ ($ \sim 300$ for a flexible ring) is a characteristic length beyond which the ring (without excluded volume) swells due to the unknotting TC~\cite{Grosberg_2000}. It has been shown that this topological swelling is characterized by the excluded volume exponent $\nu_0$.

To determine the exponent $\beta$ in Eq.~(\ref{F_unknot_scaling}), we require the squeezing free energy to be an extensive quantity; $F_{unknot}(kN, kR^3) = k F_{unknot}(N,R^3)$ for arbitrary positive number $k$. This leads to $\beta = 6$ for $N < N_0$ or $\beta = 3/(3\nu_0-1)$ for $N \gg N_0$~\cite{Sakaue_2006}. With this exponent and the definition of $X$ (Eq.~(\ref{X_define})) and $X_e$ (Eq.~(\ref{Xe_define})), the free energy~(\ref{F_unknot_scaling}) is rewritten as
\begin{eqnarray}
\frac{F_{unknot}}{k_BT} \sim \left\{
\begin{array}{ll}
\frac{N}{N_e} \left( \frac{X_e}{X}\right)^{2} & \quad (N < N_0) \\
c_0 \left( \frac{N}{N_e}\right) \left( \frac{X_e}{X}\right)^{1/(3\nu_0-1)}  & \quad (N \gg N_0)
\end{array}
\right. \label{F_unknot}
\end{eqnarray}
where a factor $c_0 = (N_0/N_e)^{3(1-2\nu_0)/[2(3\nu_0)-1]}$ of order unity arises since two characteristic topological lengths $N_0$ and $N_e$ are generally different (though their relation is not well understood yet). In all the calculation in main text, we adopted the free energy form for $N \gg N_0$ with $c_0=1$. It would be more accurate to interpolate the two expressions for $F_{unknot}$ as a function of $N$, but we have carefully checked that this does not cause any essential change in the results presented in main text~\footnote{Even if we use the first line of Eq.~(\ref{F_unknot}) instead of the second in the free energy minimization calculation, the results basically agrees with those in Fig.~\ref{Fig2} with deviation smaller than the symbol size there.}. As long as $F_{unknot}$ is written in the form $\sim (N/N_e) (X_e/X)^{1/(3\\{\tilde \nu}-1)}$, the result of free energy minimization is insensitive to the value of ${\tilde \nu}$ (either $\nu_0$ or $1/2$; the latter case corresponds to the first line of Eq.~(\ref{F_unknot}), that is $F_{unknot}$ with $N < N_0$)~\footnote{In Cates and Deutsch argument, they adopted as a competing term with $F_{unlik}$ the free energy of squeezing an ideal chain into a narrow space of size $R$  (which may be regarded as a proxy of $F_{unknot}$ here)~\cite{Cates_Deutush_1986}. This amounts to set $\beta=2$ in Eq.~(\ref{F_unknot_scaling}), which is not compatible with the extensivity requirement discussed in the text; see ref.~\cite{Sakaue_2006} for scaling structure of confining a (non-ideal) chain in a closed space.}.

\subsection*{Dependence of coordination number on the measure of ring size}
\begin{figure}[ht]
\centering
\includegraphics[width=0.38\textwidth]{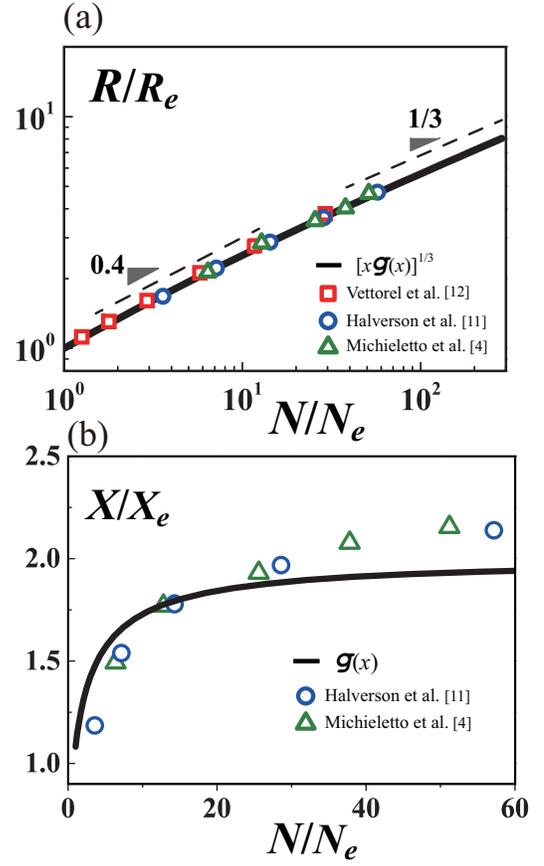}
\caption{Comparison of free energy prediction (solid line) with numerical simulation data for spatial size $R$ and coordination number $X$ of rings in their non-concatenated melt. Unlike in Fig.~2 in the main text, here the measure of ring size is the gyration radius $R=R^{(g)}$. Numerical data are obtained from literature  (squares~\cite{Vettorel_2009}, circles~\cite{Halverson_2011}, triangles~\cite{Michieletto_2016})  (a) Plot of $R/R_e$ as a function of $N/N_e$.  (b) Plot of $X/X_e$ as a function of $N/N_e$. See Table 1 in the main text for details of simulation models and the estimated values of $R_e$ and $X_e$.
}
\label{SFig1}
\end{figure}

The coordination number $X \sim R^3/N$ has an arbitrariness in its quantitative evaluation depending on how we measure the ring size $R$. In the main text, we adopt $R \equiv R^{(s)}$, where $R^{(s)}$ is the spanning distance between monomers $N/2$ apart along the ring. Here, we repeat the same analysis by adopting the gyration radius as a measure, i.e., $R \equiv R^{(g)}$. The dependence of $R^{(g)}$ on $N$ is measured in simulations~\cite{Vettorel_2009, Halverson_2011, Michieletto_2016}. These data are plotted in Fig.~\ref{SFig1} (a) along with the theoretical prediction given in the main text. Similarly, the dependence of the $R^{(g)}$-based coordination number $X|_{R= R^{(g)}}$ on $N$ is measured in simulations~\cite{Halverson_2011, Michieletto_2016}, and compared with the theoretical master curve in Fig.~\ref{SFig1} (b). The plot of Fig.~\ref{SFig1} (a) is virtually indistinguishable from that of Fig.~2 (a) in the main text. The plot of $X$ measured in simulations in Fig.~\ref{SFig1}(b) is slightly off from the free energy prediction compared with the excellent match seen in Fig.~2 (b) in the main text. This latter point may be related to the smallness of the numerical value $X|_{R= R^{(g)}}$, i.e.,$X_e|_{R= R^{(g)}} \sim 1.3 - 1.4$ (see Table~1 in the main text), indicating the sensitivity to sampling fluctuation. Yet, the deviation is at most on the order of 10 percent, and the general agreement on the $N$ dependence is still remarkable.  Therefore, we conclude that as expected from theory, the results are insensitive to the size measure upon proper rescaling. This highlights an importance of the notion of topological length scale $N_e$ and associated coordination number $X_e$ in the problem. We also note that the estimated numerical values for the rescaling factors $R_e$, $X_e$ listed in Table~1 in the main text are consistent with the assumption on the onset of the TC,i.e., the ideal chain statistics applies up to the chain length $N_e$ (see discussion in the Section {\it Effect of chain stiffness and concentration} in this Appendix).

The comparison done here indicates that, for most of practical purposes (numerical or theoretical analysis), it would be more useful to adopt the definition of $X$ in terms of $R^{(s)}$ rather than $R^{(g)}$; the numerical value is more robust against the sampling fluctuation, and fits in well with that expected from the general packing problem as discussed in the main text.

\subsection*{Onset of compact statistics}

In the main text, we have shown that the effective exponent $\nu_{eff}(N)$ for the size $R \sim N^{\nu_{eff}}$ of ring continuously decreases before reaching the asymptotic compact statistic regime with $\nu=1/3$ (Fig. 2 in the main text). Here we give a rough estimate for the ring length $N_c$, which signals the onset of the compact statistics regime. 
For convenience, we rewrite the free energy given as Eq.~(1) in main text as $F = F_{unlink}(\Psi) + F_{unknnot}(\Psi; N)$, where $F_{unlink}(\Psi)/k_BT = -\ln{[1-\Psi]}$ and $F_{unknnot}(\Psi; N) /k_BT= (N/N_e)\left( \Psi_e/\Psi \right)^{1/(3\nu_{0}-1)}$.

First, we note that the limit $\Psi ( \equiv X/X_c) \rightarrow \Psi_c \simeq 1$ signals the onset of the compact statistics. In other words, $X_c$ is the maximum coordination number, which can be achieved in dense ring polymer solutions. To deal with this limit, let us introduce a number $S \gg 1$ such that $F_{unlink}(\Psi_c) /k_BT = S$, thus $\Psi_c = 1- e^{-S}$. The free energy balance indicates $F_{unknot}(\Psi_c; N_c) /k_BT \simeq S$, which leads to $N_c/N_e \simeq \left[(1-e^{-S})/\Psi_e\right]^{1/(3\nu_{0}-1)} S \simeq \Psi_e^{1/(1-3\nu_{0})} S$.

We determine the value $S$ from the following observation. If we count the number of the pair of overlapping rings at $N = N_c$, it is evaluated as $X_c/2$ per ring. Given that $k_BT$ is the natural energy scale, it would be plausible to say that the energetic penalty of $k_BT$ per each overlap at the maximum coordination provides a proper energetic measure for $\Psi \rightarrow 1$ limit, which yields $S \simeq X_c/2$.

We thus find $N_c/N_e \sim \Psi_e^{1/(1-3\nu_{0})} X_c/2$.
With $\Psi_e \simeq 0.5$ and $X_c = 15 \sim 20$, the above argument provides the estimate $N_c/N_e \sim 30$. Here, from the consonance with the packing picture discussed in main text (last part of Sec.~\ref{TC_free_energy}), we adopt the coordination number defined in terms of the spanning distance as the measure of ring size.
Although crude, this discussion indicates a broad range for the crossover $N_c/N_e \sim 30$ (between ideal and compact statistic regimes) in agreement with the numerical simulation results.

\begin{figure}[h]
\centering
\includegraphics[width=0.3\textwidth]{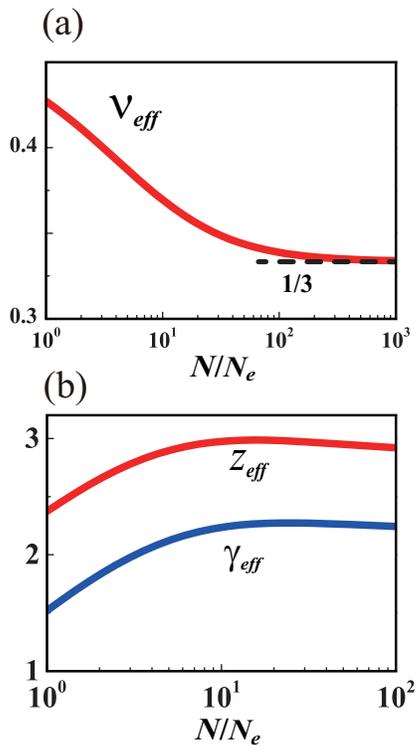}
\caption{(a) Effective size exponent $\nu_{eff}$ defined as $R(N) \sim N^{\nu_{eff}}$. (b) Effective exponents of global relaxation time $z_{eff}$ and diffusion coefficient $\gamma_{eff}$ defined, respectively, as  $\tau(N) \sim N^{z_{eff}}$ and $D(N) \sim N^{-\gamma_{eff}}$.  
}
\label{SFig2}
\end{figure}

In Fig.~\ref{SFig2}, we plot the prediction on the evolution of various effective exponents with the increase in $N$ during the crossover regime.

Note that while ideally $\nu_{eff} \to 1/2$ for $N \to N_e+0$ is expected, Fig.~\ref{SFig2} (a) shows a slightly smaller value. This may be associated with the mean-field nature of the theory. In fact, around the onset of TC $N \sim N_e$, the free energy due to TC is comparable to the thermal energy, therefore, the value determined by the free energy minimization does not necessarily correspond to the actual value.

Finally we mention that although most of recent works admit the asymptotic size exponent $\nu=1/3$, there is another claim against it~\cite{Lang_2012}. If $\nu > 1/3$ in the long chain limit, it implies the coordination number $M$ grows with $N$ without saturation, and hence the dynamical and diffusion exponents $z$ and $\gamma$ keep increasing with $N$.

\subsection*{Effect of chain stiffness and concentration}
\paragraph*{Topological length scale $N_e$}
Consider a dense solution of rings with segment volume fraction $\phi$, in which individual rings are composed of $N$ beads of size $a$. There is a bending potential introducing the directional persistence of bonds such that the segment length $l > a$.
For short rings with $N < N_e$, the ideal chain statistics implies the ring size $R \sim l M^{1/2} = a p^{1/2} N^{1/2}$, where $M = aN/l$ is the number of segments and $p = l/a$ is the segment aspect ratio. 
The coordination number is $X \sim R^3\phi/(M v_{seg}) \sim R^3 \phi/(N a^3)$, where $v_{seg} \sim a^2 l$ is the segment volume. At the onset of TC, we have $X_e \sim R_e^3 \phi/(N_e a^3)$ with $R_e \sim a p^{1/2}N_e^{1/2}$, which is converted to
\begin{eqnarray}
N_e \sim \frac{X_e^2}{\phi^2 p^3}
\label{Ne_p_phi}
\end{eqnarray}
where the proportionality constant depends on the definition of the measure of ring size ($R^{(s)}$ or $R^{(g)}$ etc.). This is very similar to the entanglement criterion in dense solution of linear polymers due to Kavassalis-Noolandi~\cite{Kavassalis_1987, Rubinstein_Colby}. The observation of constant $X_e$ (see the main text) indicates the concentration and stiffness dependence of the topological length scale $N_e \sim \phi^{-2}p^{-3}$, the systematic investigation of which in dense solution of ring polymers would be very interesting. Note that in ``genuin" semidilute regime $\phi \lesssim p^{-3}$, the correlation effect will alter the dependence as $N_e \sim \phi^{1/(1-3\nu_0)}p^{3(2-3\nu_0)/(1-3\nu_0)} \sim \phi^{-5/4}p^{-3/4}$ with the SAW exponent $\nu_0$, where the final expression is obtained by the Flory approximation $\nu_0 \simeq 3/5$.

\paragraph*{Ring size}
From the definition of the coordination number, the ring size is written as $R \sim a (NX/\phi)^{1/3}$. For long ring $N \gg N_e$, the equilibrium $X$ is determined by the free energy discussed in the main text.
Since the free energy is constructed based on the coordination number, it is independent of $\phi$ and $p$. Given this point in mind, let us rescale the ring size in unit of $R_e \sim a p^{1/2}N^{1/2}$. After some arrangement, we find
\begin{eqnarray}
\frac{R}{R_e} \sim \left( \frac{N}{N_e}\right)^{1/3} \left( \frac{X}{X_e}\right)^{1/3} = \left( \frac{N}{N_e}\right)^{1/3} \left[ g(N/N_e)\right]^{1/3}
\end{eqnarray}
where the function $g(x)$ follows from the free energy, thus, is independent of $\phi$ and $p$, too, with the asymptotic behaviors discussed in the main text. Therefore, the dependence on $\phi$ and $p$ is absorbed in rescaling factors $R_e$ and $N_e$, and one can collapse various data set with different system parameters onto the master curve as demonstrated in the main text.

In the long chain limit $N \gg N_e$, the ring size is
\begin{eqnarray}
R &\sim& R_e \left( \frac{X_c}{X_e}\right)^{1/3} \left( \frac{N}{N_e}\right)^{1/3} \nonumber \\
&\sim& a p^{1/2}N_e^{1/6}\left( \frac{X_c}{X_e}\right)^{1/3} N^{1/3}  \nonumber \\
&\sim& a \left( \frac{X_c}{\phi}\right)^{1/3} N^{1/3}
\end{eqnarray}
where Eq.~(\ref{Ne_p_phi}) is used in the last relation.

\paragraph*{Quantitative evaluation of $R_e$}
Since the ideal chain statistics without TC applies for short rings with $N < N_e$, it is possible to provide a quantitative estimate for $R_e$.
In terms of mean square size, we find
\begin{eqnarray}
(R_e^{(s)})^2 = \frac{p N_e a^2}{4} \\
(R_e^{(g)})^2 = \frac{p N_e a^2}{12}
\end{eqnarray}
for spanning size $R^{(s)}$ and the gyration radius $R^{(g)}$, respectively, at the onset of TC. With $p \simeq 2 l_p/a$ expected for worm-like chain, one can see that the values used in Fig.~2 (a) in the main text and Fig.~S1 (a), which are summarized in Table 1 in the main text, are indeed reasonable. In addition, the relation $(R_e^{(s)})^2/(R_e^{(g)})^2 = 3$ indicates the ratio $X_e|_{R= R^{(s)}}/X_e|_{R= R^{(g)}} \simeq 3^{3/2} \simeq 5.2$ between the coordination numbers based on two different measures for ring size. As seen in Table 1 in the main text, this is almost the case within statistical accuracy.
 
\bibliographystyle{apsrev4-1}
\bibliography{refs}

\end{document}